\documentclass[pdflatex,sn-vancouver-num]{sn-jnl}

\usepackage{graphicx}%
\usepackage{multirow}%
\usepackage{amsmath,amssymb,amsfonts}%
\usepackage{amsthm}%
\usepackage{mathrsfs}%
\usepackage[title]{appendix}%
\usepackage{xcolor}%
\usepackage{textcomp}%
\usepackage{manyfoot}%
\usepackage{booktabs}%
\usepackage{algorithm}%
\usepackage{algorithmicx}%
\usepackage{algpseudocode}%
\usepackage{listings}%

\theoremstyle{thmstyleone}%
\theoremstyle{thmstyletwo}%

\theoremstyle{thmstylethree}%

\begin{document}

\title[Article Title]{MAD: A Benchmark for Multi-Turn Audio Dialogue Fact-Checking}



\author*[1]{\fnm{Chaewan} \sur{Chun}}\email{czc5884@psu.edu}

\author[1]{\fnm{Lysandre} \sur{Terrisse}}\email{lft5243@psu.edu}

\author[1]{\fnm{Delvin Ce} \sur{Zhang}}\email{delvincezhang@gmail.com}

\author[1]{\fnm{Dongwon} \sur{Lee}}\email{dongwon@psu.edu}

\affil*[1]{\orgname{The Pennsylvania State University}, \orgaddress{\city{University Park}, \postcode{16802}, \state{PA}, \country{USA}}}


\abstract{Despite the growing popularity of audio platforms, fact-checking spoken content remains significantly underdeveloped. Misinformation in speech often unfolds across multi-turn dialogues, shaped by speaker interactions, disfluencies, overlapping speech, and emotional tone—factors that complicate both claim detection and verification. Existing datasets fall short by focusing on isolated sentences or text transcripts, without modeling the conversational and acoustic complexity of spoken misinformation. We introduce MAD (Multi-turn Audio Dialogues), the first fact-checking dataset aligned with multi-turn spoken dialogues and corresponding audio. MAD captures how misinformation is introduced, contested, and reinforced through natural conversation. Each dialogue includes annotations for speaker turns, dialogue scenarios, information spread styles, sentence-level check-worthiness, and both sentence- and dialogue-level veracity. The dataset supports two core tasks: check-worthy claim detection and claim verification. Benchmarking shows that even strong pretrained models reach only 72-74\% accuracy at the sentence level and 71-72\% at the dialogue level in claim verification, underscoring MAD’s difficulty. MAD offers a high-quality benchmark for advancing multimodal and conversational fact-checking, while also surfacing open challenges related to reasoning over speech and dialogue dynamics.
}




\maketitle

\section{Introduction}

\begin{figure}[ht]
  \centering
  \begin{minipage}[b]{0.43\textwidth}
    \centering
    \includegraphics[width=\linewidth]{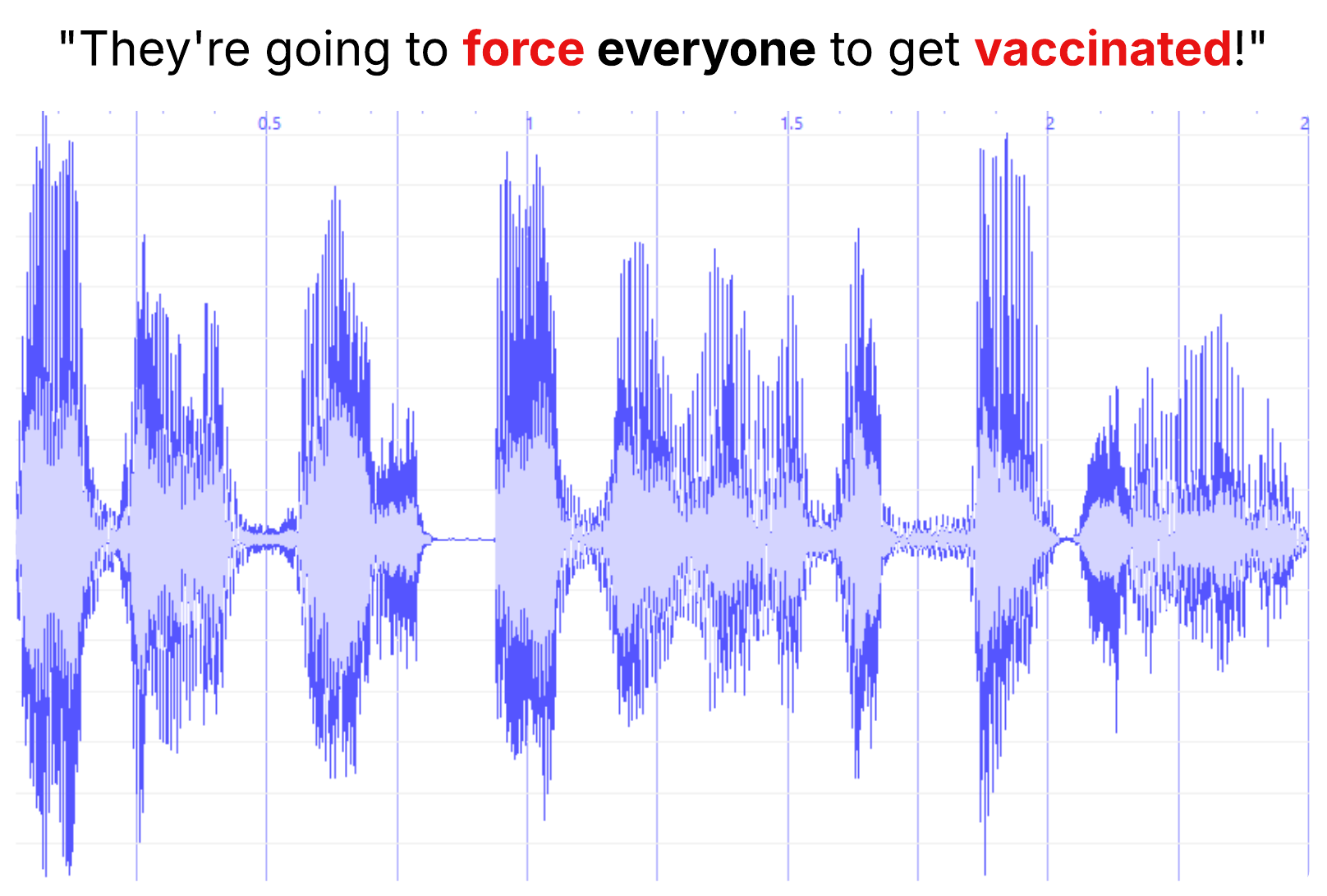}
    \caption{Audio waveform analysis on one of the check-worthy claims in MAD.}
    \label{fig:1}
  \end{minipage}\hfill
  \begin{minipage}[b]{0.54\textwidth}
    \centering
    \includegraphics[width=\linewidth]{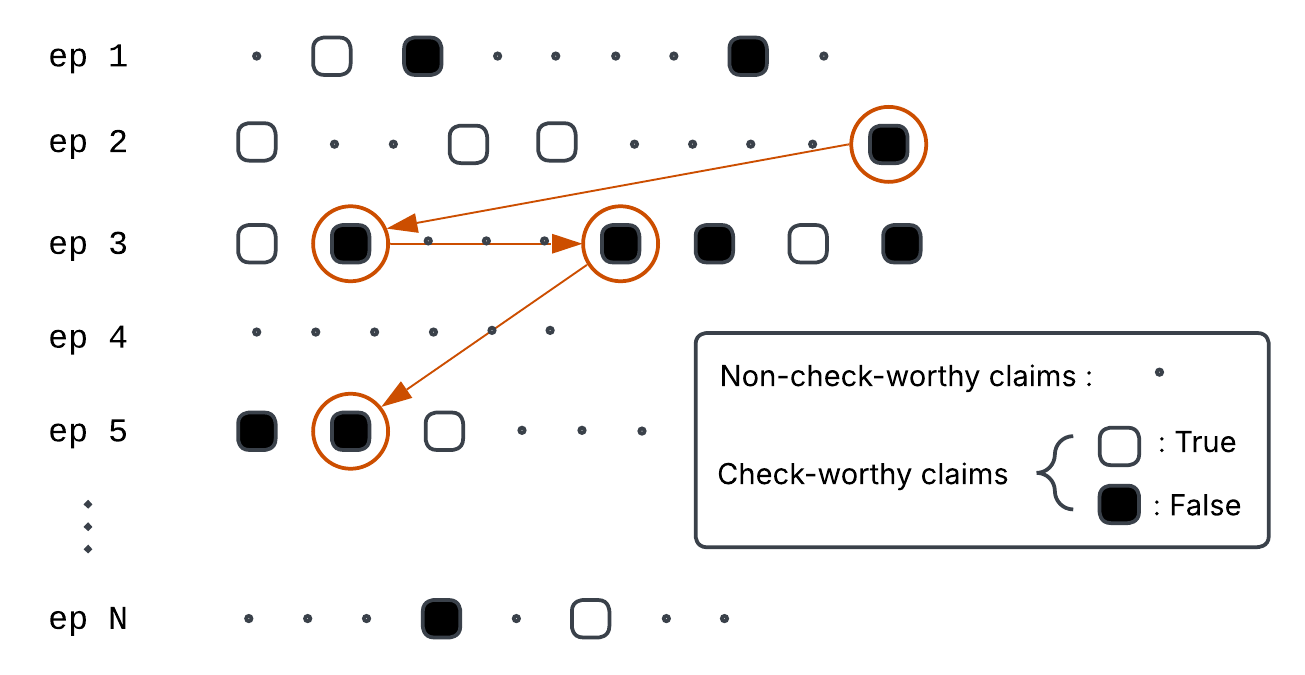}
    \caption{Illustration of claim detection and verification tasks.
    }
    \label{fig:2}
  \end{minipage}
\end{figure}

Spoken claims rarely stand alone — they build up across conversational turns. In spoken media such as podcasts, radio shows, and live discussions, misinformation often unfolds gradually, not through standalone sentences but through layered, interactive exchanges. A seemingly neutral utterance may take on misleading significance only when interpreted alongside earlier and/or later conversational turns. Claims may be implied, collaboratively constructed, or reinforced across multiple speakers, which challenges systems that treat utterances in isolation. Yet despite the growing influence of spoken content, audio remains significantly underexplored in fact-checking research, especially compared to text and image modalities.

These challenges stem not only from dialogue structure but from the inherent complexities of spoken language. Conversations are spontaneous, filled with disfluencies, overlapping speech, and co-references. Intonation and emotional tone shape how messages are perceived, and boundaries between claims and commentary are often blurred. As a result, existing models developed for structured, well-formed text may struggle to generalize to this fluid, multi-speaker environment. Moreover, audio carries prosodic and paralinguistic cues, such as emotional tone, emphasis, hesitation, or sarcasm, which are often lost in transcripts but can profoundly affect how claims are interpreted. These subtleties, such as rising pitch or elongation at key moments as shown in Figure~\ref{fig:1}, may signal uncertainty, exaggeration, or intent, providing critical context for fact-checking models.

Progress in audio fact‑checking research is constrained by the absence of dialogue‑structured, multimodal datasets with comprehensive verification annotations. While audio deepfake detection has benefited from numerous benchmarks, resources for verifying the factuality of spoken claims remain scarce. For instance, \citet{ctfcc18} offers short, isolated statements from political speeches with veracity labels, but omits multi‑turn conversational structure. Similarly, \citet{livefc} processes segmented claims from live audio with speaker diarization, yet does not model interactive conversations. \citet{settybecker} annotates claim check‑worthiness in podcasts, but again, without any multi-turn dialogue framework. \citet{dialfact} introduces a well‑structured dialogue dataset with veracity labels, yet lacks sentence‑level check‑worthiness annotations and does not include any audio data. Consequently, none of the existing audio datasets integrates multi‑turn dialogue, check‑worthy claim annotations, claim‑level veracity labels, and aligned audio data. On the contrary, our dataset contains all four types of data and is the first to support end‑to‑end audio dialogue fact‑checking.

We present MAD (Multi-turn Audio Dialogues) as an initial step toward multimodal fact-checking in dialogue-centric audio content. Rather than focusing on isolated claims, MAD captures how misinformation may emerge and evolve through interaction. While modest in scale, it spans diverse spreading scenarios, speaker roles, and rhetorical strategies. The dataset supports two key fact-checking tasks: 1) check-worthy claim detection, which identifies utterances that merit verification, and 2) claim verification, which assesses the veracity of those claims. We benchmarked two pretrained encoders (RoBERTa-base \cite{roberta}, DeBERTa-v3-base \cite{debertav3}) and Llama 3(8B) \cite{llama3paper} fine-tuned with QLoRA \cite{qlora} for the claim verification task on MAD. The highest performing model among those three achieved around 74\% accuracy at sentence level, with performance dropping to about 72\% at the dialogue level, showing the difficulty of the MAD dataset. This resource offers a foundation for studying fact-checking in conversational audio and highlights the challenges and opportunities for future work in this space, particularly within the contexts of \textit{Social Cyber Security} and \textit{Trust in AI} within the SBP-BRiMS.

\section{Dataset Creation}

Most misinformation datasets are constructed by collecting content from online platforms or news sources. In contrast, MAD (Multi-turn Audio Dialogues) takes a generative approach. We create synthetic but realistic dialogues centered around fact-checked political claims. This design enables controlled exploration of misinformation dynamics in conversational settings.

\subsection{Data Collection}
To construct our dialogue-based benchmark dataset, we began with the LIAR dataset \cite{liar}, an established benchmark for fake news detection comprising over 12,800 manually labeled short political statements from PolitiFact\footnote{\url{https://www.politifact.com}}. Specifically, we randomly selected 600 claims, evenly split between 300 labeled ``true'' and 300 labeled ``false''.
For dialogue generation, we utilized Gemini 2.5 Pro \cite{gemini25}, which offers an exceptionally large context window, allowing us to include rich structured prompts and maintain coherence across long dialogues. We designed these prompts to produce natural, diverse English-language conversations based on each LIAR claim. Each dialogue centers on a single claim and inherits its veracity label (True or False) from the LIAR dataset.

The dialogues were crafted to reflect diverse interpersonal dynamics and rhetorical strategies, enabling the study of how factual claims are introduced, debated, reinforced, or refuted in everyday conversation. To maximize variation and minimize systematic bias, the generation process followed a structured three-part configuration:

\paragraph{Part A: Fictional Speaker Profiles}
Each dialogue features two fictional speakers defined by randomly assigned attributes, including their relationship (e.g., coworkers, siblings, in-laws, old friends), a core personality trait (e.g., anxious, cynical, gullible, enthusiastic), and relevant background interests (e.g., politically engaged, avoids the news, fascinated by history). These attributes shaped each speaker’s tone, perspective, and rhetorical tendencies, enhancing realism and speaker diversity.

\paragraph{Part B: Claim-Introduction Styles}

We further propose three methods by which a claim is formed and spread.

\begin{itemize}
    \item \textbf{Anecdote-Driven Framing:} Emphasizing personal or social consequences through storytelling.
    \item \textbf{Sarcastic Framing:} Using irony or ridicule to challenge opposing viewpoints.
    \item \textbf{Tentative Rumor Framing:} Informally introducing the claim with hedging language (e.g., ``I heard...'', ``Some say...).
\end{itemize}

\paragraph{Part C: Dialogue Scenarios}

To make the dataset diverse and challenging, we explore five dialogue scenarios about how a claim is discussed and conveyed.

\begin{itemize}
    \item \textbf{Informer vs. Skeptic:} One speaker asserts the claim confidently, while the other challenges or critiques it.
    \item \textbf{Believer vs. Believer:} Both speakers affirm the claim, reinforcing it through mutual agreement.
    \item \textbf{Doubter vs. Doubter:} Shared skepticism and joint critical inquiry shape the interaction.
    \item \textbf{Presenter vs. Deflector:} A potential debate fizzles into evasion, ambiguity, or unresolved tension.
    \item \textbf{Convincer vs. Inquirer:} One speaker advocates strongly, while the other probes critically without immediate dismissal.
\end{itemize}

Dialogues were generated in twelve batches of 50 dialogues (six batches per veracity label), with Parts A–C sampled randomly for each instance. To ensure high linguistic and contextual quality, each batch underwent a secondary quality screening using GPT-4o \cite{gpt4o}. Specifically, the generated dialogues were evaluated for naturalness, coherence, contextual appropriateness, and avoidance of repetitive phrasing across batches. Batches that failed to meet these quality standards were regenerated and re-evaluated until they received a satisfactory assessment.

Additionally, each sentence in the dialogues was annotated with a binary ``check-worthy'' label, indicating whether it substantively referenced, questioned, or elaborated on the underlying claim. These annotations support a fine-grained check-worthy claim detection task, while the claim-level veracity label enables claim verification task. Together, these annotations and dataset features make the MAD benchmark particularly well-suited for multi-level misinformation analysis in audio dialogue contexts.

\subsection{Dataset Statistics}

\begin{table}[t]
\centering
\caption{Overall dataset statistics for the 600-dialogue benchmark.}
\label{tab:dataset-stats}
\begin{tabular}{@{}lr@{}}
\toprule
\textbf{Statistic} & \textbf{Value} \\
\midrule
Total dialogues                              & 600 \\
Claim veracity (False / True)                & 300 / 300 \\
Total sentences                              & 4,915 \\
Check-worthy sentences                       & 1,748 (35.56\%) \\
\midrule
Turns per dialogue (min / max / mean ± SD)   & 4 / 6 / 4.45 ± 0.61 \\
Sentences per dialogue (min / max / mean ± SD) & 5 / 15 / 8.19 ± 2.23 \\
Tokens per sentence (min / max / mean ± SD)  & 1 / 46 / 12.22 ± 7.32 \\
Tokens per dialogue (min / max / mean ± SD)  & 59 / 152 / 100.06 ± 19.32 \\
\bottomrule
\end{tabular}
\end{table}

\begin{table}[t]
\centering
\caption{Distribution of claim-introduction styles and dialogue scenarios.
}
\label{tab:scenario-spread}
\begin{tabular}{@{}lrr@{}}
\toprule
\textbf{Category}                          & \textbf{Type} & \textbf{Count} \\ 
\midrule
\multirow{3}{*}{Part B: Claim-Introduction Styles}
                    & Anecdote-Driven Framing                    & 226 \\
                    & Sarcastic Framing                              & 178 \\
                    & Tentative Rumor Framing                        & 196 \\
\midrule                               
\multirow{5}{*}{Part C: Dialogue Scenarios}      
                                & Informer vs.\ Skeptic        & 136 \\
                                & Believer vs.\ Believer                       & 140 \\
                                & Doubter vs.\ Doubter                      &  86 \\
                               & Presenter vs.\ Deflector                       & 108 \\
                               & Convincer vs.\ Inquirer              & 130 \\
\bottomrule
\end{tabular}
\end{table}


Table~\ref{tab:dataset-stats} reports that MAD comprises 600 dialogues totaling 4,915 sentences, with 35.6\% labeled as check‑worthy. Dialogues average 8.2 sentences and 100 tokens, spanning 4–6 turns. Table~\ref{tab:scenario-spread} shows balanced coverage across three claim introduction styles and five dialogue scenarios, supporting both sentence- and dialogue-level misinformation tasks.

To illustrate how MAD captures conversational misinformation dynamics, we present an example dialogue centered on a fact-checked political claim. The example reflects different veracity labels, dialogue scenarios, and information spread styles. As shown, sentences labeled as check-worthy (marked [CW] after sentence) highlight moments where claims are asserted, questioned, or elaborated.



\paragraph{Claim (False):} A recent drop in the unemployment rate is questionable because it was calculated over two days of the Thanksgiving week. \vspace{0.1in}\\
\textbf{Dialogue Scenario:} Presenter vs. Deflector \\
\textbf{Dialogue Style:} Sarcastic Framing

\begin{quote}
\textbf{SPEAKER\_0:} You can't trust any of the good economic news they put out. Rush Limbaugh pointed out that a recent drop in the unemployment rate was totally questionable because they calculated it over just two days during Thanksgiving week. It's a sham.\textsuperscript{[CW]}\\
\textbf{SPEAKER\_1:} Dad, that's not how they calculate the unemployment rate. It's a massive, month-long survey of households. They don't just measure it for two days.\textsuperscript{[CW]}\\
\textbf{SPEAKER\_0:} Oh, so the government's 'official' story is the truth? Come on. They cooked the books. They picked two days during a holiday week to make it look good. It's obvious.\textsuperscript{[CW]}\\
\textbf{SPEAKER\_1:} That's a conspiracy theory. The Bureau of Labor Statistics has been using the same methodology for decades, through Republican and Democratic administrations. It's not a political operation.\\
\textbf{SPEAKER\_0:} Fine, believe the swamp if you want to.
\end{quote}

\subsection{Audio Generation}
To support multimodal research, MAD benchmark includes synthesized audio for all 600 dialogues using XTTS-v2 \cite{xtts}, a multilingual text-to-speech (TTS) model capable of expressive, speaker-distinct synthesis. For each dialogue, two synthetic voices were randomly selected from XTTS-v2's pool of pretrained speaker embeddings. Turns were synthesized independently and stitched together to produce coherent, natural-sounding conversations.

While our experiments in this work are text-based, the inclusion of audio enables future research directions that text alone cannot capture. Spoken content carries rich paralinguistic cues, such as stress, pacing, hesitation, and sarcasm, that shape how claims are framed and received. These features influence perceived credibility and emotional salience, yet are often lost in transcription \cite{ivanov, whatsapp}. For example, hesitation may indicate uncertainty, while exaggerated pitch or repetition may signal persuasive emphasis.
Importantly, spoken misinformation often gains traction through conversational interaction. Repeated agreement, passive listening, or even neutral responses can function as implicit endorsements, amplifying misinformation across turns. This multi-speaker amplification effect plays a key role in real-world misinformation dynamics.

\section{Experiments and Results}\label{sec3}
We evaluate our dataset on two tasks: 1) \textit{check-worthy claim detection}, which identifies which conversational utterances merit fact-checking attention, and 2) \textit{claim verification}, which determines the veracity of claims labeled as check-worthy, as shown in Figure~\ref{fig:2}. We adopt standard supervised classification setups for both tasks, fine-tuning transformer-based language models. We follow existing works \cite{dialfact, dialogue, correct} and report mean and standard deviation of accuracy and F1 scores across three random seeds.

\subsection{Check-Worthy Claim Detection}

We fine-tuned RoBERTa-base \cite{roberta} and DeBERTa-v3-base \cite{debertav3} as binary classifiers to distinguish check-worthy from non-check-worthy utterances. We trained each model to classify whether a claim is check-worthy, using the claim and its dialogue context as input features. Inputs were tokenized using HuggingFace’s AutoTokenizer with a maximum sequence length of 512 tokens and padded to the nearest multiple of 8. Both models used a hidden and classifier dropout of 0.2 for regularization.

To prevent data leakage, we performed train/test splits at the dialogue level using group identifiers from the dialogue metadata. To address class imbalance, we applied per-seed undersampling to balance positive and negative instances in training. Models were fine-tuned for 3 epochs with a batch size of 8, learning rate of 2e\textsuperscript{-5}, weight decay of 0.05, and warm-up ratio of 0.1. We evaluated and checkpointed every 60 steps, applying early stopping based on validation F1 score (patience = 3). The two models shared the same training procedure, differing only in backbone architecture and tokenization. As shown in Table~\ref{tab:checkworthy}, both RoBERTa‑base and DeBERTa‑v3‑base achieved stable performance with around 86\% accuracy and 82\% F1.

\begin{table}[t]
\centering
\caption{Check-worthy claim detection performance (mean ± SD across three seeds).}
\label{tab:checkworthy}
\begin{tabular}{@{}lcc@{}}
\toprule
\textbf{Metric} & \textbf{RoBERTa-base} & \textbf{DeBERTa-v3-base} \\
\midrule
Accuracy  & 86.95 ± 1.00 & 86.36 ± 1.73 \\
F1 Score  & 82.71 ± 1.34 & 81.94 ± 1.53 \\
\bottomrule
\end{tabular}
\end{table}

\subsection{Claim Verification}

We evaluate claim verification using RoBERTa-base \cite{roberta}, DeBERTa-v3-base \cite{debertav3}, and Meta-Llama 3 (8B) \cite{llama3paper}, modeling spoken misinformation at both the sentence and dialogue levels.

\textbf{Sentence-Level Classification.} For RoBERTa and DeBERTa, we fine-tune pretrained models using Hugging Face Transformers on a binary classification task. Each input consists of a check-worthy claim and its surrounding dialogue context, tokenized to a maximum length of 512 tokens. Train/test splits are stratified by dialogue groups to prevent leakage across samples from the same conversation. We train for 2 epochs using a batch size of 8, learning rate of 1.5e\textsuperscript{-5}, weight decay of 0.1, and warm-up ratio of 0.1. Early stopping is triggered based on validation F1.

Llama 3 is fine-tuned using QLoRA, which applies 4-bit quantization and LoRA adapters to the attention projection layers (\texttt{q\_proj}, \texttt{k\_proj}, \texttt{v\_proj}, \texttt{o\_proj}). We train for 4 epochs with a batch size of 1 and gradient accumulation of 8 steps, using a learning rate of 1e\textsuperscript{-5} and a dropout rate of 0.1. Early stopping is triggered after 2 non-improving epochs based on validation loss. After training, the LoRA adapters are merged into the base model for inference. Table~\ref{tab:ver} shows that all three models achieve comparable performance, with encoder-only models slightly leading.

\textbf{Dialogue-Level Aggregation.}  
For all models, we extract sentence-level [CLS] embeddings and predicted probabilities from the fine-tuned sentence classifiers. These are used as input features to a hierarchical dialogue-level classifier.

For RoBERTa and DeBERTa, we use a lightweight Transformer-based aggregator with 2 layers and 2–4 attention heads. Sentence embeddings and probabilities are concatenated and passed through a Transformer encoder with soft attention pooling, followed by a binary classification head. The aggregator is trained using AdamW (learning rate = 1e\textsuperscript{-4}), class-weighted loss, and early stopping based on validation F1 with a 20\% held-out split. Training is capped at 50 epochs.

In contrast, the Llama aggregator incorporates additional regularization and architectural customization. Sentence-level [CLS] embeddings are z-normalized using the global mean and variance, then projected separately from predicted probabilities. These vectors are combined with learned positional encodings and passed through a Transformer encoder layer with soft attention pooling. The model is trained in two stages: first updating only the classification head, then fine-tuning the full model with a reduced learning rate. R-Drop regularization is applied by performing two forward passes per input and minimizing the sum of cross-entropy loss and symmetric KL divergence. Gradients are clipped to a maximum norm of 1.0, early stopping is applied based on validation F1 using a 20\% held-out split, and class priors are encoded into the classifier bias to improve stability. Table~\ref{tab:ver} shows that RoBERTa-base and DeBERTa-v3-base maintain strong performance from sentence to dialogue level, with only marginal changes, indicating stable generalization. Llama 3 with QLoRA performance drops more noticeably, highlighting the increased difficulty of aggregating across dialogue and the limitations of lightweight aggregation on LLM-derived representations.

\begin{table}[t]
\centering
\caption{Claim verification performance at sentence and dialogue levels (mean ± SD, three seeds).}
\label{tab:ver}
\begin{tabular}{@{}llccc@{}}
\toprule
\textbf{Level} & \textbf{Metric} & \textbf{RoBERTa‑base} & \textbf{DeBERTa‑v3‑base} & \textbf{Llama3 (8B)+QLoRA} \\
\midrule
Sentence‑level & Accuracy & 72.60±4.04 & 74.42±3.29 & 70.62±3.73 \\
               & F1 Score & 68.69±7.67 & 73.28±2.65 & 68.50±6.45 \\
\midrule
Dialogue‑level & Accuracy & 72.50±2.45 & 71.67±2.04 & 58.33±5.07 \\
               & F1 Score & 71.45±1.79 & 71.05±1.08 & 59.80±5.55 \\
\bottomrule
\end{tabular}
\end{table}

\section{Discussion and Future Work}

While MAD represents an important initial step toward enabling fact-checking research on dialogue-based audio content, several limitations remain. Addressing these will be key to expanding its utility and real-world relevance.

\textbf{Synthetic Dialogue vs. Transcript-Based Verification.} In this paper, we did not use audio transcripts. Instead, we relied on generated conversational dialogue to simulate multi-turn misinformation. While this enabled controlled claim construction, it omits key features such as disfluencies, overlapping speech, and prosodic cues. In future work, we plan to incorporate aligned audio recordings and transcripts to support more authentic spoken-language fact-checking.

\textbf{Multilingual and Source Expansion.} The current dataset is limited to English and based solely on short political claims from the LIAR dataset \cite{liar}, limiting both linguistic and topical diversity. Since misinformation is a global issue spanning multiple domains, we plan to expand by incorporating multilingual dialogues and sourcing claims from diverse origins, including structured sources (e.g., fact-checking archives) and unstructured content (e.g., podcasts, interviews). This will broaden coverage beyond politics and enable more realistic modeling of misinformation.

\textbf{Long-Form Dialogue Scaling.} Our dialogues are currently short and self-contained, whereas real-world misinformation often unfolds across long-form content. We plan to include full podcast episodes or talk-show segments with aligned audio to support tasks like temporal reasoning, coreference resolution, and claim tracking across turns.

\textbf{Dialogue Authenticity \& Speaker Metadata.} Future dataset versions will include real-world dialogues from public audio platforms, annotated with speaker diarization, demographic metadata, and rhetorical styles to capture richer interpersonal dynamics.

\textbf{Audio Realism \& Evaluation.} We aim to enhance the quality of synthetic audio to better mimic human prosody, emotion, and rhythm. Beyond standard text-to-speech, we will explore expressive voice models and validate output through human perception studies evaluating naturalness and believability.

\textbf{New Tasks: Evidence Retrieval \& Explanation.} In future versions, we plan to support additional tasks such as evidence retrieval and explanation generation by linking claims to external sources and providing human-readable justifications. This will foster fact-checking systems that not only flag questionable content but also explain their reasoning.

\textbf{Dataset Scaling.} We aim to expand the dataset to thousands of dialogues across diverse topics, speakers, and interaction lengths. Scaling both volume and variety will support more robust and generalizable models for real-world fact-checking.


\section{Conclusion}

Our work introduces MAD, the first benchmark to combine multi-turn dialogues, sentence-level check-worthiness, claim-level veracity, and aligned audio for spoken misinformation research. MAD captures how false claims emerge and evolve through conversation, addressing key gaps in existing datasets. Baseline experiments show that while transformer models perform well on check-worthy claim detection, claim verification, particularly at the dialogue level, remains challenging. By integrating dialogue structure and audio cues, MAD provides a foundation for future research on conversational misinformation and prosody-aware fact-checking.


\bibliography{sn-bibliography}

\end{document}